\documentclass[12pt]{iopart}

\usepackage{graphicx}% Include figure files
\usepackage{dcolumn}% Align table columns on decimal point
\usepackage{bm}% bold math
\usepackage{epsfig}

\begin{document}

\title[Thermodynamics and spin-charge separation of
1D repulsive 3-component fermions]{Thermodynamics and spin-charge
separation of one-dimensional strongly repulsive three-component
fermions}

\author{Peng He$^{1,2}$, Jen Yee Lee$^{2}$, Xiwen Guan$^{2}$, Murray
T. Batchelor$^{2,3}$ and Yupeng Wang$^{1}$}

\address{$^{1}$ Beijing National Laboratory for Condensed Matter
Physics, Institute of Physics, Chinese Academy of Sciences, Beijing
100190, P. R. China}%

\address{$^{2}$ Department of Theoretical Physics, Research
School of Physics and Engineering, Australian National University,
Canberra ACT 0200, Australia}

\address{$^{3}$ Mathematical Sciences Institute, Australian
National University, Canberra ACT 0200, Australia}

\ead{hepeng@iphy.ac.cn}

\begin{abstract}
The low temperature thermodynamics of one-dimensional strongly
repulsive $SU(3)$ fermions in the presence of a magnetic field is
investigated via the Yang-Yang thermodynamic Bethe ansatz method.
The analytical free energy and magnetic properties of the model at
low temperatures in a weak magnetic field are derived via the
Wiener-Hopf method. It is shown that the low energy physics can be
described by spin-charge separated conformal field theories of an
effective Tomonaga-Luttinger liquid and an antiferromagnetic $SU(3)$
Heisenberg spin chain. Beyond the Tomonaga-Luttinger liquid regime,
the equation of state is given in terms of the polylog function for
a weak external field. The results obtained are essential for
further study of quantum criticality in strongly repulsive
three-component fermions.

\end{abstract}

%Uncomment for PACS numbers title message
% Keywords required only for MST, PB, PMB, PM, JOA, JOB?
%\vspace{2pc}
%\noindent{\it Keywords}: Article preparation, IOP journals
% Uncomment for Submitted to journal title message
\submitto{\JPA}
% Comment out if separate title page not required
\maketitle

\section{Introduction}

Recent experiments on quantum gases in one-dimension (1D)
\cite{TGexp, Toshiya, Moritz05, Haller, Hulet} provide an exciting
way to test 1D many-body physics extensively studied in the
literature via the Bethe ansatz (BA) \cite{Takahashi}, effective
field theory \cite{Giamarchi-b} and other theoretical methods. Novel
1D many-body physics  such as the phenomena of spin-charge
separation \cite{Recati,Fuchs,Kollath}, universal Luttinger liquid
thermodynamics \cite{Blote,Affleck,Guan,Erhai,JYLee} and quantum
criticality, are quite different from higher dimensional physics. 1D
many-body systems also provide insights into higher dimensional
physics. For example, theoretical predictions for the existence of a
Fulde-Ferrell-Larkin-Ovchinnikov (FFLO) \cite{FFLO} like pairing
state in the 1D interacting Fermi gas have emerged
\cite{Guan,Orso,Hu,Wadati,Mueller,
Ueda,Feiguin,Gao,Rizzi2008,Batrouni2008,Kinnunen2006,Zhao2008}.

A scheme for mapping out the physical properties of homogeneous
systems by the inhomogeneity of the trap has been successfully
applied to ultracold atom experiments \cite{Hulet}, since the
thermodynamics of interacting fermionic systems can be measured
explicitly and the interaction between fermions can be tuned
precisely via a broad Feshbach resonance.  In particular, fermionic
alkaline-earth atoms display an exact $SU(N)$ spin symmetry with
$N=2I+1$ where $I$ is the nuclear spin \cite{Gorshkov}. For example,
a recent experiment dramatically realized the model of fermionic
atoms  with  $SU(2)\times SU(6) $ symmetry where electron spin
decouples from  its  nuclear spin  $I=5/2$ for ${}^{171}$Yb
\cite{Taie}. Such fermionic systems with enlarged $SU(N)$ spin
symmetry are expected to display a remarkable diversity of new
quantum phases and quantum critical phenomena  due to the existence
of multiple charge bound states \cite{Ho,Ho2}. The strongly
attractive multi-component Fermi gases with higher spin symmetries
have also been investigated via the Bethe ansatz method
\cite{Guan2008prl,Guan2,He}. These developments provide an exciting
opportunity to explore the universal thermodynamics and quantum
critical behavior of strongly interacting fermions with high spin
symmetries in 1D.

The Bethe ansatz has proven to be a powerful method to study 1D
quantum gases. This method was applied to the 1D Bose gas by Lieb
and Liniger \cite{Lieb} and to the spin-1/2 Fermi gas by Yang
\cite{Yang1967} and Gaudin \cite{Gaudin1967}. After that, the
multi-component Fermi gas was studied by Sutherland
\cite{Sutherland}. The study of the thermodynamics of the attractive
Fermi gas was initiated by Yang \cite{Yang-a} and Takahashi
\cite{Takahashi-2}. Moreover, the thermodynamics of 1D Fermi gases
is also widely studied for its rich physics. For spin-$1/2$ fermions
with attractive interaction, the phase diagrams and Luttinger liquid
physics have been investigated by using both analytical
\cite{Guan,Erhai} and numerical methods \cite{Hu,Orso}. The key
features of this $T=0$ phase diagram were experimentally confirmed
using finite temperature density profiles of trapped fermionic
${}^6$Li atoms \cite{Hulet}.

For the strongly attractive spin-1/2 Fermi gas at finite
temperatures, the thermodynamics of the homogeneous system is
described by two coupled Fermi gases of bound pairs and excess
fermions in the charge sector and ferromagnetic  spin-spin
interaction in the spin sector \cite{Guan,Erhai}. Spin fluctuations
are suppressed by a strong effective magnetic field  at low
temperatures. However, for the repulsive case, the situation is
quite different since spin fluctuations play a dominant role in the
low energy physics. Spin fluctuations are described by the $SU(2)$
antiferromagnetic spin-spin Heisenberg system   and thus the
coupling of charge and spin parts make the thermodynamic Bethe
ansatz (TBA) equations more complex.  So far we know that the
Wiener-Hopf method can be applied here to deal with the spin part of
the TBA equations for weak magnetic field. This method was used by
Mezincescu \textit{et al.} to investigate the thermodynamics of
Heisenberg spin chains with $SU(2)$ and $SU(3)$ symmetries
\cite{mezincescu, mezincescu2}. Recently, Lee \textit{et al.}
\cite{JYLee} applied the Wiener-Hopf method to spin-$1/2$ repulsive
fermions with $SU(2)$ symmetry and studied the low temperature
thermodynamics and correlation functions via the Bethe ansatz method
analytically. The Wiener-Hopf technique was applied to get  the
solutions of the TBA equations in the spin sector. Then Sommerfeld's
expansion was utilized in the low temperature limit with strong
interactions to get the multi-component Tomonaga-Luttinger liquid
(TLL) form of the free energy, which shows a spin-charge separation
with the central charge of the spin and charge parts both equal to
$1$.

Spin-charge separation is a hallmark of 1D many-body physics. It is
a universal feature that interacting particles ``split'' into spins
and charges as temperature tends to absolute zero temperature. The
collective excitations with only spin or charge are called spinon
and chargon/holon (the antiparticle of chargon), which have
different velocities. This behavior was studied in different kinds
of materials \cite{Recati,Fuchs,Kollath} and recently observed in
experiments, for example, of 1D metallic wires on surfaces, 1D
organic wires, carbon-nanotubes, quantum wires in semiconductors,
and other types of 1D systems. 1D quantum systems with high spin
symmetries will give diverse magnetism and exotic properties for
these kinds of gases. However, such systems are more difficult to
solve \cite{Tsvelik,Zhang,Wang} due to the complicated magnetic
ordering. The three-component Fermi gas has $U(1)\times SU(3)$
symmetry which leads to two sets of spin waves.  It is very
interesting to see  how the low temperature thermodynamics  of such
a gas naturally separates into free Gaussian field theories for the
$U(1)$ charge degree of freedom and the two spin sectors.

In this paper, 1D fermions with $U(1)\times SU(3)$ symmetry is
studied by using the Wiener-Hopf method.  The result shows that the
system with strong repulsive interactions at low temperatures has
the central charge $C_{c}=1$ for the charge sector and central
charge $C_{s}=2$ for the two spin sectors. We also derive the
leading order finite temperature corrections of the free energy
which is consistent with expectations from conformal field theory
\cite{Blote,Affleck, Centralcharge, Kawakami}. In addition, the
polylog function is used to derive the equation of state in a wider
temperature regime.

This paper is organized as follows. In Section II,  the   BA and TBA
equations are presented. In Section III, the Wiener-Hopf method is
applied to solve the TBA equations for the spin sectors. In Section
IV, the thermodynamics of the system is derived via Sommerfeld's
expansion for low temperatures and strong interactions. We also show
that the expression for the free energy at low temperatures is in
agreement with conformal field theory and derive the spin and charge
velocities. Section V is reserved for  the conclusion and
discussion. Some details are given in the appendices.

\section{The Model and thermodynamic Bethe ansatz}
Consider a 1D system of $N$ fermions with mass $m$ and a spin
independent $\delta$-function potential. The fermions can occupy
three possible hyperfine levels ($\left\vert 1\right\rangle $,
$\left\vert 2\right\rangle $, and $\left\vert 3\right\rangle $) with
particle numbers $ N^{1}$, $N^{2}$, and $N^{3}$, respectively. They
are also constrained to a line of length $L$ with periodic boundary
conditions. This system is described by the many-body Hamiltonian
\cite{Sutherland,Takahashi-2}
\begin{equation}
\mathcal{H}_0=-\frac{\hbar ^{2}}{2m}\sum_{i=1}^{N}\frac{\partial
^{2}}{
\partial x_{i}^{2}}+g_{1D}\sum_{1\leq i<j\leq N}\delta
(x_{i}-x_{j})+E_z. \label{Ham}
\end{equation}
Here $E_z=\sum_{i=1}^3 N^i \epsilon^i_Z$ is the Zeeman energy. The
spin independent contact interaction $g_{1D}$ exists between
fermions with different hyperfine states so that the number of
fermions in each spin state is conserved. It is positive for
repulsive interaction and negative for attractive interaction. For
simplicity, we set $\hbar=2m=1$. They can be reintroduced when
necessary. It is possible to tune scattering lengths between atoms
in different sublevels to form nearly $SU(3)$ degenerate Fermi gases
via a broad Feshbach resonance.

The Hamiltonian in equation (\ref{Ham}) exhibits a symmetry of
$U(1)\times SU(3)$, where $U(1)$ and $SU(3)$ are the symmetries for
charge and spin degrees of freedom, respectively. As mentioned
earlier, this model was solved via a nested BA \cite{Sutherland,
Takahashi-2}. The energy eigenspectrum is given by $E=\frac{\hbar
^2}{2m}\sum_{j=1}^Nk_j^2$ in terms of the quasimomenta
$\left\{k_j\right\}$ satisfying the BA equations
\cite{Sutherland,Takahashi-2}
\begin{eqnarray}
e^{\mathrm{i}k_{j}L}&=&\prod_{\ell=1}^{M_{1}}\frac{k_{j}-\Lambda
_{\ell}+\mathrm{i}c/2}{k_{j}-\Lambda _{\ell}-\mathrm{i}c/2},
\nonumber
\\
\prod_{j=1}^{N}\frac{\Lambda _{\ell}-k_{j}+\mathrm{i}c/2}{\Lambda
_{\ell}-k_{j}-\mathrm{i}c/2}&=&-\prod_{\alpha
=1}^{M_{1}}\frac{\Lambda _{\ell}-\Lambda _{\alpha
}+\mathrm{i}c}{\Lambda _{\ell}-\Lambda _{\alpha }-\mathrm{i}c}
\prod_{m=1}^{M_{2}}\frac{\Lambda _{\ell}-\lambda
_{m}-\mathrm{i}c/2}{\Lambda _{\ell}-\lambda
_{m}+\mathrm{i}c/2}, \nonumber\\
 \prod_{\ell=1}^{M_{1}}\frac{\lambda
_{m}-\Lambda _{\ell}+\mathrm{i}c/2}{\lambda _{m}-\Lambda
_{\ell}-\mathrm{i}c/2}&=&-\prod_{\beta =1}^{M_{2}}\frac{\lambda
_{m}-\lambda _{\beta }+\mathrm{i}c}{\lambda _{m}-\lambda _{\beta
}-\mathrm{i}c}. \label{BAE}
\end{eqnarray}
Here $j=1,\ldots ,N$, $\ell =1,\ldots ,M_{1}$ and $m=1,\ldots
,M_{2}$ with quantum numbers $M_{1}=N^{2}+N^{3}$, and $M_{2}=N^{3}$.
We define the interaction strength as $c=m g_{1D}/\hbar^2>0$ since
we only consider the repulsive case. The parameters
$\left\{k_{j}\right\}$ are quasimomenta and $\left\{\Lambda
_{\ell},\lambda_{m}\right\} $ are the rapidities that characterize
the internal hyperfine spin degrees of freedom. In the thermodynamic
limit, $N, L \rightarrow \infty$ with the ratio for linear particle
density $n=N/L$ kept finite. Notice that there are two classes of
strings. Both sets of solutions can be expressed in terms of the
string hypothesis
\begin{eqnarray}
\Lambda _{j}^{n,\alpha} &=& \Lambda
_{j}^{n}-\frac{1}{2}(n+1-2\alpha)\mathrm{i}|c|,\qquad \alpha = 1,\ldots,n \\
\lambda _{j}^{n,\beta} &=& \lambda
_{j}^{n}-\frac{1}{2}(n+1-2\beta)\mathrm{i}|c|,\qquad \beta =
1,\ldots,n
\end{eqnarray}
where $n$ is the length of each string, $j$ labels each individual
string, and $\Lambda _{j}^{n}$ and $\lambda _{j}^{n}$ are the real
parts of the $\Lambda$ and $\lambda$ strings, respectively. To
distinguish the rapidities in different spin levels, define
$\lambda^{(r)}$ as
\begin{eqnarray}
\Lambda:=\lambda^{(1)}, \qquad \lambda:=\lambda^{(2)}.
\end{eqnarray}

For repulsive interaction, the quasimomenta $\{k_j\}$ are real,
while $\{\lambda^{(r)}\}$ form complex spin strings which
characterize the spin wave fluctuations at finite temperatures.
Without loss of generality, we consider the case where equal Zeeman
splitting occurs between fermions of different spin to simplify our
analysis, i.e., $H_{1}=H_{2}=H$. The equilibrium states are then
determined by minimizing the Gibbs free energy, which gives rise to
a set of coupled nonlinear integral equations -- the TBA equations,
which are
\begin{eqnarray}
\nonumber \varepsilon(k) &=& k^{2}-\mu -H
-T\sum_{n=1}^{\infty}a_{n}\ast \ln
(1+e^{-{\phi^{(1)}_n(k)}/{T}}),  \\
\nonumber\phi^{(1)}_{n}(\lambda^{(1)}) &=& n H - T a_{n}\ast \ln
(1+e^{-{\varepsilon(\lambda^{(1)})}/{T}}) + T\sum_{m}T_{mn}\ast \ln
(1+e^{-{\phi^{(1)}_n(\lambda^{(1)} )}/{T}}) \\
&& \nonumber -T \sum_{m}S_{mn}\ast \ln (1+e^{-{\phi^{(2)}_n(\lambda^{(1)})}/{T}}), \\
\phi^{(2)}_{n}(\lambda^{(2)}) &=&n H + T \sum_{m}T_{mn}\ast
\ln(1+e^{-{\phi^{(2)}_n(\lambda^{(2)})}/{T}}) -T \sum_{m}S_{mn}\ast
\ln (1+e^{-{\phi^{(1)}_n(\lambda^{(2)})}/{T}}).\label{TBA}
\end{eqnarray}
Here $\varepsilon(k)$ denotes the dressed energy and
$\phi^{(r)}_{n}$ ($r=1,2$) are associated with densities of strings
with length-$n$ in $\lambda^{(r)}$ parameter spaces. $\ast$
represents the convolution integral, i.e. $(a\ast b)(x)=\int
a(x-y)b(y)dy$, and the function $a_{m}(x)$ is given by
\begin{equation}
a_{m}(x) =\frac{1}{2\pi }\frac{mc}{(mc/2) ^{2}+x^{2}}.
\end{equation}

The functions $T_{mn}$ and $ S_{mn}$ are defined as
\begin{eqnarray}
T_{mn}(x) &=& \left\{
                \begin{array}{ll}
                  a_{m+n}(x)+2a_{m+n-2}(x)+\ldots+2a_{|m-n|+2}(x)+a_{|m-n|}(x), & \hbox{for $m\neq n$;} \\
                  2a_{2}(x)+2a_{4}(x)+\ldots+2a_{2n-2}(x)+a_{2n}(x), & \hbox{for $m=n$.}
                \end{array}
              \right. \\
S_{mn}(x) &=& \left\{
             \begin{array}{ll}
               a_{m+n-1}(x)+a_{m+n-3}(x)+\ldots+a_{|m-n|+3}(x)+a_{|m-n|+1}(x), & \hbox{for $m\neq n$;} \\
               a_{1}(x)+a_{3}(x)+\ldots+a_{2n-3}(x)+a_{2n-1}(x), & \hbox{for $m=n$.}
             \end{array}
           \right.
\end{eqnarray}
An alternative set of TBA equations are
\begin{eqnarray}
\nonumber \varepsilon(k) &=& k^{2}-\mu -T s_1 \ast a_{1}\ast \ln
(1+e^{-{\varepsilon\left(k\right)}/{T}}) \\ \nonumber && -Ts_1\ast
\ln
(1+e^{{\phi^{(1)}_1\left(k\right)}/{T}}) -Ts_2\ast \ln (1+e^{{\phi^{(2)}_1(k)}/{T}}) , \nonumber  \\
\nonumber \phi^{(1)}_{1}(\lambda^{(1)} ) &=& T s\ast \ln
\left(1+e^{{\phi^{(1)}_2(\lambda^{(1)} )}/{T}}\right)-T s\ast \ln
\left(1+e^{-{\varepsilon(\lambda^{(1)})}/{T}}\right) \\ && \nonumber
-T s \ast
\ln \left(1+e^{-{\phi^{(2)}_1(\lambda^{(1)} )}/{T}}\right),\nonumber \\
\phi^{(1)}_n(\lambda^{(1)}) &=& Ts \ast
\ln\left(1+e^{\phi^{(1)}_{n-1}(\lambda^{(1)})}\right)+Ts
\ast \ln\left(1+e^{\phi^{(1)}_{n+1}(\lambda^{(1)})}\right),\nonumber\\
\phi^{(2)}_{1}(\lambda^{(2)} ) &=& Ts \ast \ln
\left(1+e^{{\phi^{(2)}_2(\lambda^{(2)} )}/{T}}\right) -T s\ast \ln
\left(1+e^{-{\phi^{(1)}_1(\lambda^{(2)} )}/{T}}\right),\nonumber\\
\phi^{(2)}_{n}(\lambda^{(2)} ) &=& Ts \ast
\ln\left(1+e^{\phi^{(2)}_{n-1}(\lambda^{(2)})}\right)+Ts \ast
\ln\left(1+e^{\phi^{(2)}_{n+1}(\lambda^{(2)})}\right).\label{TBA2}
\end{eqnarray}
with the limit
\begin{eqnarray}
\lim_{n\rightarrow \infty} \frac{\phi^{(r)}_n ( \lambda^{(r)} )}{n}
&=&H\qquad (r=1,2) \label{limit}
\end{eqnarray}
and
\begin{eqnarray}
s(x)&=& \frac{1}{2 c \cosh \left(\pi x /c \right)},\\
s_1(x)&=& \frac{1}{\sqrt 3 c}\frac{\cosh \left(\pi x/ 3 c\right)}{\cosh \left(\pi x/c \right)},\label{s1}\\
s_2(x)&=& \frac{1}{\sqrt 3 c}\frac{\sinh \left(\pi x/ 3
c\right)}{\sinh \left(\pi x/c \right)}.\label{s2}
\end{eqnarray}
The two sets of TBA equations are interchangeable via Fourier
transformation.

\section{Wiener-Hopf Solution For The Spin Part}
The string part is the most difficult part of the TBA equations to
solve. It consists of an infinite number of string functions
$\phi^{(r)}_n(\lambda)$ which are related through the set of
equations (\ref{TBA}) or (\ref{TBA2}). These coupled nonlinear
integral equations cannot be solved analytically in most cases.
However, in some special cases the equations can be simplified. When
$T$ is very low ($T\ll 1$), these equations reduce to a set of
linearly coupled equations. Moreover, when $c$ is very large ($c\gg
1$), Taylor expansion can be applied to solve this set of equations.
In the following, we will consider the case where $T\ll 1$ and $c\gg
1$.

Observe from the second set of TBA equations (\ref{TBA2}) that
$\phi^{(r)}_n>0$ for $n>1$ at low temperatures. The functions
$s(\lambda)$ and $\ln\left(1+e^{\phi^{(r)} _n(\lambda)/T}\right)$
are always greater than zero in their entire domain, so the
convolution is always greater than zero. This positivity condition
implies that the function
$T\ln\left(1+e^{-\phi^{(r)}_n(\lambda)/T}\right)\rightarrow 0$ for
$T\rightarrow 0$. Thus all the spin string functions could be
neglected except the $n=1$ function $\phi^{(r)} _1 (\lambda)$ in the
first set of TBA equations. For $T\ll 1$, only the lowest strings
are therefore left in the strong coupling limit $c \gg 1$, and we
can rewrite the term as
\begin{equation}
T a_1 \ast \ln\left(1+e^{-\varepsilon(k)/T}\right) \approx 2\pi P
a_1(k),
\end{equation}
where
\begin{equation}
P=\frac{T}{2\pi }\int_{-\infty}^{\infty} dk\ln
\left(1+e^{-{\varepsilon \left( k\right)
}/{T}}\right).\label{pressure}
\end{equation}

The first set of TBA equations (\ref{TBA}) can thus be simplified as
\begin{eqnarray}
\varepsilon\left(k\right) &=&k^{2}-\mu -H
-T a_{1}\ast \ln (1+e^{-{\phi^{(1)}_1 (k )}/{T}}),  \\
\phi^{(1)}_{1} (\lambda) &=& H - 2\pi P a_{1}(\lambda) + T a_{2}\ast
\ln \left(1+e^{-{\phi^{(1)}_1 (\lambda)}/{T}}\right)- T a_{1}\ast
\ln \left(1+e^{-{\phi^{(2)}_1
(\lambda)}/{T}}\right), \\
\phi^{(2)}_{1}(\lambda) &=& H + T a_{2}\ast \ln
(1+e^{-{\phi^{(2)}_1(\lambda)}/{T}})-T a_{1}\ast \ln
(1+e^{-{\phi^{(1)}_1(\lambda)}/{T}}).
\end{eqnarray}
where we have switched to a common spin variable $\lambda$ to
represent both spin spaces $\lambda^{(1)}$ and $\lambda^{(2)}$.
Taking the limit $T\rightarrow 0$ yields
\begin{eqnarray}
\varepsilon\left(k\right) &=&k^{2}-\mu -H
+a_{1}\ast \phi^{(1)-}_1\left(k\right),\label{epsilon1}   \\
\phi^{(1)}_{1}(\lambda) &=& H - 2\pi P a_{1}(k)- a_{2}\ast
\phi^{(1)-}_1(\lambda) + a_{1}\ast
\phi^{(2)-}_1(\lambda), \label{phi1}\\
\phi^{(2)}_{1}(\lambda) &=& H - a_{2}\ast \phi^{(2)-}_1(\lambda) +
a_{1}\ast \phi^{(1)-}_1(\lambda).\label{phi2}
\end{eqnarray}

We have decomposed the functions as
\begin{eqnarray}
\phi^{(r)}_1(\lambda)&=&\phi^{(r)+}_1(\lambda)+\phi^{(r)-}_1(\lambda),
\end{eqnarray}
where
\begin{eqnarray}
\phi^{(r)+}_1(\lambda) &=& \left\{
                             \begin{array}{ll}
                               \phi^{(r)}_1(\lambda), & \hbox{for $\phi^{(r)}_1(\lambda)>0$;} \\
                               0, & \hbox{for $\phi^{(r)}_1(\lambda)<0$.}
                             \end{array}
                           \right. \\
\phi^{(r)-}_1(\lambda) &=& \left\{
                             \begin{array}{ll}
                               0, & \hbox{for $\phi^{(r)}_1(\lambda)>0$;} \\
                               \phi^{(r)}_1(\lambda), & \hbox{for $\phi^{(r)}_1(\lambda)<0$.}
                             \end{array}
                           \right.
\end{eqnarray}

After taking the Fourier transfors and going through some
manipulation, equations (\ref{phi1}) and (\ref{phi2}) can be written
as
\begin{eqnarray}
\phi^{(1)}_1(\lambda) &=& H - 2\pi P s_1(\lambda) +h
\ast\phi^{(1)+}_1(\lambda) +
g\ast\phi^{(2)+}_1(\lambda),\nonumber\\
\phi^{(2)}_1(\lambda) &=& H - 2\pi P s_2(\lambda)
+h\ast\phi^{(2)+}_1(\lambda) +
g\ast\phi^{(1)+}_1(\lambda),\label{phiplus}
\end{eqnarray}
where
\begin{eqnarray}
h(\lambda) &=& s_1\ast a_1(\lambda)- s_2(\lambda),\\
g(\lambda) &=& s_2\ast a_1(\lambda)- s_1(\lambda).
\end{eqnarray}
Both equations in (\ref{phiplus}) are the $T\rightarrow0$ limit
cases of
\begin{eqnarray}
\phi^{(1)}_1(\lambda) &=& H - 2\pi P s_1(\lambda) +h \ast
T\ln(1+e^{{\phi^{(1)}_1}(\lambda)/{T}}) +
g\ast T\ln(1+e^{{\phi^{(2)}_1(\lambda)}/{T}}),\nonumber\\
\phi^{(2)}_1(\lambda) &=& H - 2\pi P s_2(\lambda) +h \ast
T\ln(1+e^{{\phi^{(2)}_1}(\lambda)/{T}}) + g\ast
T\ln(1+e^{{\phi^{(1)}_1}(\lambda)/{T}}).\label{phito0}
\end{eqnarray}

In order to find a relationship between the dressed energies and the
temperature, we write $\phi^{(1)}_1(\lambda)$ as
\begin{equation}
\phi^{(r)}_1(\lambda)=\phi^{(r)}(\lambda)+\eta^{(r)}(\lambda),
\end{equation}
where the first term $\phi^{(r)}(\lambda)$ is the term of zeroth
order when $T=0$ and second term $\eta^{(r)}(\lambda)$ is the first
order correction to the limit $T\rightarrow 0$. Substituting this
equation into (\ref{phito0}) and using the expressions in
(\ref{phiplus}) for $\phi^{(r)}(\lambda)$, the equation for
$\eta^{(r)}(\lambda)$ becomes
\begin{eqnarray}
\nonumber \eta^{(1)}(\lambda)&=& h\ast \left[
T\ln\left(1+e^{(\phi^{(1)}(\lambda)+\eta^{(1)}(\lambda))/T}\right)-\phi^{(1)+}(\lambda)\right]
\\ && + g\ast \left[ T\ln \left(1+e^{(\phi^{(2)}(\lambda)
+\eta^{(2)}(\lambda))/T}\right)
-\phi^{(2)+}(\lambda) \right],\\
\nonumber \eta^{(2)}(\lambda)&=& h\ast \left[
T\ln\left(1+e^{(\phi^{(2)}(\lambda)+\eta^{(2)}(\lambda))/T}\right)-\phi^{(2)+}(\lambda)\right]
\\ && + g\ast \left[ T\ln \left(1+e^{(\phi^{(1)}(\lambda)
+\eta^{(1)}(\lambda))/T}\right) -\phi^{(1)+}(\lambda) \right].
\end{eqnarray}

Since $\phi^{(r)}(\lambda)$ is an even function and $\lambda=
\lambda^{(r)}_0$ give the zero points, the above equations for
$\eta^{(r)}(\lambda)$  can be simplified to become
\begin{eqnarray}
\nonumber
\eta^{(1)}(\lambda)&\approx&\int_{|\lambda|>\lambda_{0}^{(1)}}
h(\lambda-\lambda^\prime) \eta^{(1)}(\lambda^\prime) d
\lambda^\prime \\ && +\int_{|\lambda|>\lambda_{0}^{(2)}}
h(\lambda-\lambda^\prime)
\eta^{(2)}(\lambda^\prime)d\lambda^\prime + E_h^{(1)}(\lambda) + E_g^{(2)}(\lambda), \\
\nonumber
\eta^{(2)}(\lambda)&\approx&\int_{|\lambda|>\lambda_{0}^{(2)}}
h(\lambda-\lambda^\prime) \eta^{(2)}(\lambda^\prime)d\lambda^\prime
\\ && +\int_{|\lambda|>\lambda_{0}^{(1)}} h(\lambda-\lambda^\prime)
\eta^{(1)}(\lambda^\prime)d\lambda^\prime + E_h^{(2)}(\lambda) +
E_g^{(1)}(\lambda),\label{etaaa}
\end{eqnarray}
where the approximation
\begin{eqnarray}
&&f\ast \left[ T \ln\left(
1+e^{(\phi^{(r)}(\lambda)+\eta^{(r)}(\lambda))/T} \right)
-\phi^{(1)+}(\lambda) \right]\nonumber\\
&&= \int_{|\lambda|>\lambda_{0}^{(r)}} d\lambda^\prime
f(\lambda-\lambda^\prime) \left[ T \ln\left(
1+e^{(\phi^{(r)}(\lambda)+\eta^{(r)}(\lambda))/T} \right)
-\phi^{(1)}(\lambda)
\right]\nonumber\\
&&
\quad+\int_{-\lambda_0^{(r)}}^{\lambda_0^{(r)}}f(\lambda-\lambda^\prime)
T \ln\left( 1+e^{(\phi^{(r)}(\lambda)+\eta^{(r)}(\lambda))/T} \right)\nonumber\\
&&\approx \int_{|\lambda|>\lambda_{0}^{(r)}} d\lambda^\prime
f(\lambda-\lambda^\prime) \eta^{(r)}(\lambda^\prime) +
E_f^{(r)}(\lambda).
\end{eqnarray}
has been made. Here $E_f^{(r)}(\lambda)$ stands for the integral
\begin{eqnarray}
E_f^{(r)}(\lambda):=
\int_{-\infty}^{\infty}f(\lambda-\lambda^\prime) T \ln\left(
1+e^{|\phi^{(r)}(\lambda)|/T} \right)
\end{eqnarray}
and $f$ represents the subscripts $h$ and $g$. For $T\rightarrow 0$,
the major contribution to the integral is from the regions near the
zero points of $\phi^{(r)}(\lambda)$, i.e.
$\lambda=\pm\lambda^{(r)}_0$. Therefore, the expansion of
$\phi^{(r)}(\lambda)$ around $\lambda=\pm\lambda^{(r)}_0$ is
\begin{eqnarray}
\phi^{(r)}(\lambda) = t^{(r)}\left( \lambda - \lambda^{(r)}_0
\right) + O\left[ \left( \lambda - \lambda^{(r)}_0 \right)^2 \right]
\end{eqnarray}
where
\begin{eqnarray}
t^{(r)}:=\frac{d \phi^{(r)}} {d\lambda}
\bigg|_{\lambda=\lambda^{(r)}_0}
\end{eqnarray}
Then the leading term of $E_f^{(r)}(\lambda)$ becomes
\begin{eqnarray}
E_f^{(r)}(\lambda)&\approx& \frac{2T^2}{t^{(r)}}\left[f\left(\lambda
- \lambda_0^{(r)}\right) + f\left(\lambda+ \lambda_0^{(r)} \right)
\right]\int_0^\infty d x \ln\left(1+e^{-x}\right)\nonumber\\
&=&\frac{\pi^2 T^2}{6 t^{(r)}}\left[f\left(\lambda -
\lambda_0^{(r)}\right) + f\left(\lambda+ \lambda_0^{(r)} \right)
\right].
\end{eqnarray}

Substituting these results into ({\ref{etaaa}}), we obtain the
equations for $\phi^{(r)}$ and $\eta^{(r)}$, namely
\begin{eqnarray}
\phi^{(1)}(\lambda) &=& H - 2\pi P s_1(\lambda) +h
\ast\phi^{(1)+}(\lambda) +
g\ast\phi^{(2)+}(\lambda),\label{phi11}\\
\phi^{(2)}(\lambda) &=& H - 2\pi P s_2(\lambda)
+h\ast\phi^{(2)+}(\lambda) +
g\ast\phi^{(1)+}(\lambda),\label{phi12}\\
\eta^{(1)}(\lambda)&\approx& \int_{|\lambda|>\lambda_{0}^{(1)}}
h(\lambda-\lambda^\prime) \eta^{(1)}(\lambda^\prime) d
\lambda^\prime +\int_{|\lambda|>\lambda_{0}^{(1)}}
h(\lambda-\lambda^\prime)
\eta^{(2)}(\lambda^\prime)d\lambda^\prime\nonumber\\
&& \nonumber +\frac{\pi^2 T^2}{6 t^{(1)}}\left[h\left(\lambda -
\lambda_0^{(1)}\right) + h\left(\lambda+ \lambda_0^{(1)} \right)
\right] \\ && + \frac{\pi^2 T^2}{6 t^{(2)}}\left[g\left(\lambda -
\lambda_0^{(1)}\right) + g\left(\lambda+ \lambda_0^{(1)} \right)
\right], \label{eta11}\\
\eta^{(2)}(\lambda)&\approx&\int_{|\lambda|>\lambda_{0}^{(2)}}
h(\lambda-\lambda^\prime) \eta^{(2)}(\lambda^\prime)d\lambda^\prime
+\int_{|\lambda|>\lambda_{0}^{(2)}} h(\lambda-\lambda^\prime)
\eta^{(1)}(\lambda^\prime)d\lambda^\prime \nonumber\\
&& \nonumber +\frac{\pi^2 T^2}{6 t^{(2)}}\left[h\left(\lambda -
\lambda_0^{(2)}\right) + h\left(\lambda+ \lambda_0^{(2)} \right)
\right] \\ && + \frac{\pi^2 T^2}{6 t^{(1)}}\left[g\left(\lambda -
\lambda_0^{(2)}\right) + g\left(\lambda+ \lambda_0^{(2)} \right)
\right].\label{eta12}
\end{eqnarray}
When $H\rightarrow 0$, $\lambda_0^{(r)}\rightarrow \infty$,
(\ref{phi11}) and (\ref{phi12}) can be simplified to
\begin{equation}
\phi^{(r)}(\lambda) = H - 2\pi P s_r(\lambda).
\end{equation}
Since $\phi^{(r)}(\lambda^{(r)}_0)=0$, $s_r(\lambda^{(r)}_0)$ can be
written as
\begin{equation}
s_r(\lambda^{(r)}_0)=\frac{H}{2\pi P}.\label{zeroorder1}
\end{equation}
From (\ref{s1}) and (\ref{s2}) we find that in the limit
$\lambda_{0}^{(r)}\gg 1$,
\begin{equation}
s_{r}(\lambda^{(r)}_0)=\frac{1}{\sqrt{3}c}e^{{-2\pi
\lambda^{(r)}_0}/{3c}}+O\left(\frac{1}{c^2}\right).\label{zeroorder2}
\end{equation}
Hence from both equations (\ref{zeroorder1}) and (\ref{zeroorder2})
we obtain
\begin{eqnarray}
\lambda_0^{(r)}=-\frac{3c
}{2\pi}\left[\ln\left(\frac{\sqrt{3}c}{2\pi}H\right)+\ln
\kappa^{(r)}\right]
\end{eqnarray}
where $\kappa^{(r)}\approx 1$ is an integral constant.

Instead of $\phi^{(r)}(\lambda)$ and $\eta^{(r)}(\lambda)$, it is
easier to work with the functions
\begin{eqnarray}
S^{(r)}(\lambda) &=& \left\{
                       \begin{array}{ll}
                         e^{{2\pi \lambda_0^{(r)}}/{3c}}\kappa^{(r)}\phi^{(r)}(\lambda+\lambda^{(r)}_0), & \hbox{for $\lambda>0$;} \\
                         0, & \hbox{for $\lambda<0$.}
                       \end{array}
                     \right. \\
T^{(r)}(\lambda) &=& \left\{
                       \begin{array}{ll}
                         \frac{6e^{{-2\pi \lambda_0^{(r)}}/{3c}}}{\pi^2T^2\kappa^{(r)}} \eta^{(r)}(\lambda+\lambda^{(r)}_0), & \hbox{for $\lambda>0$;} \\
                         0, & \hbox{for $\lambda<0$.}
                       \end{array}
                     \right.
\end{eqnarray}

Notice that $\phi^{(r)}(\lambda)>0$ for
$|\lambda|>\lambda_{0}^{(r)}$. Instead, we shift the integration
variables so that they run from $0$ to $\infty$. Observe that
$h(\lambda+2\lambda_{0}^{(r)})$ and $g(\lambda +\lambda_{0}^{(1)}
+\lambda_{0}^{(2)})$ vanish as $H\rightarrow 0$ ($\lambda_{0}^{(r)}
\rightarrow \infty$) for finite $\lambda$. The functions $g(\lambda
+\lambda_{0}^{(1)} -\lambda_{0}^{(2)})$ and $g(\lambda
-\lambda_{0}^{(1)} +\lambda_{0}^{(2)})$ remain finite. Equations
(\ref{phi11})--(\ref{eta12}) in the form of the newly introduced
functions are
\begin{eqnarray}
\nonumber S^{(1)}(\lambda)&=& \frac{2\pi
P}{\sqrt{3}c}\left(1-\kappa^{(1)}e^{-2\pi \lambda/3c}\right)
+\int_0^\infty h(\lambda-\lambda^\prime) S^{(1)} (\lambda^\prime)
d\lambda^\prime \\ && + \int_0^\infty g(\lambda+\lambda_0^{(1)}
-\lambda^\prime-\lambda_0^{(2)}) S^{(2)} (\lambda^\prime)
d\lambda^\prime, \\
\nonumber T^{(1)}(\lambda)&=& \frac{h(\lambda)}{S^{(1)\prime}(0)} +
\frac{g(\lambda+\lambda^{(1)}_0-\lambda^{(2)}_0)}{S^{(2)\prime}(0)}
+\int_0^\infty h(\lambda-\lambda^\prime) T^{(1)} (\lambda^\prime)
d\lambda^\prime \\ && + \int_0^\infty g(\lambda+\lambda_0^{(1)}
-\lambda^\prime-\lambda_0^{(2)}) T^{(2)} (\lambda^\prime)
d\lambda^\prime, \\
\nonumber S^{(2)}(\lambda)&=& \frac{2\pi
P}{\sqrt{3}c}\left(1-\kappa^{(2)}e^{-2\pi \lambda/3c}\right)
+\int_0^\infty h(\lambda-\lambda^\prime) S^{(2)} (\lambda^\prime)
d\lambda^\prime \\ && + \int_0^\infty g(\lambda+\lambda_0^{(2)}
-\lambda^\prime-\lambda_0^{(1)}) S^{(1)} (\lambda^\prime)
d\lambda^\prime, \\
\nonumber T^{(2)}(\lambda)&=& \frac{h(\lambda)}{S^{(2)\prime}(0)} +
\frac{g(\lambda+\lambda^{(2)}_0-\lambda^{(1)}_0)}{S^{(1)\prime}(0)}
+\int_0^\infty h(\lambda-\lambda^\prime) T^{(2)} (\lambda^\prime)
d\lambda^\prime \\ && + \int_0^\infty g(\lambda+\lambda_0^{(2)}
-\lambda^\prime-\lambda_0^{(1)}) T^{(1)} (\lambda^\prime)
d\lambda^\prime.
\end{eqnarray}
Writing the equations in standard Wiener-Hopf form for
$-\infty<\lambda<\infty$ gives
\begin{eqnarray}
\nonumber S^{(1)}(\lambda)&=& f^{(1)}_S(\lambda) +b^{(1)}_S(\lambda)
+\int_{-\infty}^\infty h(\lambda-\lambda^\prime) S^{(1)}
(\lambda^\prime) d\lambda^\prime \\ && + \int_{-\infty}^\infty
g(\lambda+\lambda_0^{(1)} -\lambda^\prime-\lambda_0^{(2)}) S^{(2)}
(\lambda^\prime) d\lambda^\prime, \label{WHS1}\\
\nonumber T^{(1)}(\lambda)&=& f^{(1)}_T(\lambda) +b^{(1)}_T(\lambda)
+\int_{-\infty}^\infty h(\lambda-\lambda^\prime) T^{(1)}
(\lambda^\prime) d\lambda^\prime \\ && + \int_{-\infty}^\infty
g(\lambda+\lambda_0^{(1)} -\lambda^\prime-\lambda_0^{(2)}) T^{(2)}
(\lambda^\prime) d\lambda^\prime, \label{WHT1}\\
\nonumber S^{(2)}(\lambda)&=& f^{(2)}_S(\lambda) +b^{(2)}_S(\lambda)
+\int_{-\infty}^\infty h(\lambda-\lambda^\prime) S^{(2)}
(\lambda^\prime) d\lambda^\prime \\ && + \int_{-\infty}^\infty
g(\lambda+\lambda_0^{(2)} -\lambda^\prime-\lambda_0^{(1)}) S^{(1)}
(\lambda^\prime) d\lambda^\prime, \label{WHS2}\\
\nonumber T^{(2)}(\lambda)&=& f^{(2)}_T(\lambda) +b^{(2)}_T(\lambda)
+\int_{-\infty}^\infty h(\lambda-\lambda^\prime) T^{(2)}
(\lambda^\prime) d\lambda^\prime \\ && + \int_{-\infty}^\infty
g(\lambda+\lambda_0^{(2)} -\lambda^\prime-\lambda_0^{(1)}) T^{(1)}
(\lambda^\prime) d\lambda^\prime, \label{WHT2}
\end{eqnarray}
where
\begin{eqnarray}
f^{(r)}_{S}(\lambda) &=& \left\{
                         \begin{array}{ll}
                           \frac{2\pi P}{\sqrt{3}c}\left(1-\kappa^{(r)}e^{{-2\pi \lambda}/{3c}}\right), & \hbox{for $\lambda>0$;} \\
                           0, & \hbox{for $\lambda<0$.}
                         \end{array}
                       \right. \\
b^{(1)}_{S}(\lambda) &=& \left\{
                         \begin{array}{ll}
                           0, & \hbox{for $\lambda>0$;} \\
                           -h*S^{(1)}(\lambda)-g*S^{(2)}(\lambda+\lambda_0^{(1)}-\lambda_0^{(2)}), & \hbox{for $\lambda<0$.}
                         \end{array}
                       \right. \\
b^{(2)}_{S}(\lambda) &=& \left\{
                         \begin{array}{ll}
                           0, & \hbox{for $\lambda>0$;} \\
                           -h*S^{(2)}(\lambda)-g*S^{(1)}(\lambda+\lambda_0^{(1)}-\lambda_0^{(2)}), & \hbox{for $\lambda<0$.}
                         \end{array}
                       \right. \\
f^{(1)}_{T}(\lambda) &=& \left\{
                         \begin{array}{ll}
                           \frac{h(\lambda)}{S^{(1)\prime}(0)}+\frac{g(\lambda+\lambda^{(1)}_0-\lambda^{(2)}_0)}{S^{(2)\prime}(0)}, & \hbox{for $\lambda>0$;} \\
                           0, & \hbox{for $\lambda<0$.}
                         \end{array}
                       \right. \\
f^{(2)}_{T}(\lambda) &=& \left\{
                         \begin{array}{ll}
                           \frac{h(\lambda)}{S^{(2)\prime}(0)}+\frac{g(\lambda+\lambda^{(1)}_0-\lambda^{(2)}_0)}{S^{(1)\prime}(0)}, & \hbox{for $\lambda>0$;} \\
                           0, & \hbox{for $\lambda<0$.}
                         \end{array}
                       \right. \\
b^{(1)}_{T}(\lambda) &=& \left\{
                         \begin{array}{ll}
                           0, & \hbox{for $\lambda>0$;} \\
                           -h*T^{(1)}(\lambda)-g*T^{(2)}(\lambda+\lambda_0^{(1)}-\lambda_0^{(2)}), & \hbox{for $\lambda<0$.}
                         \end{array}
                       \right. \\
b^{(2)}_{T}(\lambda) &=& \left\{
                         \begin{array}{ll}
                           0, & \hbox{for $\lambda>0$;} \\
                           -h*T^{(2)}(\lambda)-g*T^{(1)}(\lambda+\lambda_0^{(1)}-\lambda_0^{(2)}), & \hbox{for $\lambda<0$.}
                         \end{array}
                       \right.
\end{eqnarray}
The equations (\ref{WHS1})-(\ref{WHT2}) can be solved via Fourier
transform. Define the Fourier coefficients of $S^{(r)}(\lambda)$ and
$T^{(r)}(\lambda)$ as
\begin{eqnarray}
\hat{S}^{(r)}(\omega)&=&\int_{-\infty}^\infty d\lambda
e^{\mathrm{i}\lambda
\omega} S^{(r)}(\lambda),\\
\hat{T}^{(r)}(\omega)&=&\int_{-\infty}^\infty d\lambda
e^{\mathrm{i}\lambda \omega} T^{(r)}(\lambda).
\end{eqnarray}
The functions $\hat{S}^{(r)}(\omega)$ and $\hat{T}^{(r)}(\omega)$
are analytic on the upper-half-plane.

For simplicity, we write equations (\ref{WHS1})--(\ref{WHT2}) in
matrix form
\begin{eqnarray}
\mathbf{S}(\lambda)&=& \mathbf{f_S}(\lambda)+\mathbf{b_S}(\lambda)
+\int_{-\infty}^{\infty}
\mathbf{K}(\lambda-\lambda^\prime)\mathbf{S}(\lambda^\prime)d\lambda',\\
\mathbf{T}(\lambda)&=& \mathbf{f_T}(\lambda)+\mathbf{b_T}(\lambda)
+\int_{-\infty}^\infty
\mathbf{K}(\lambda-\lambda^\prime)\mathbf{T}(\lambda^\prime)
d\lambda',
\end{eqnarray}
where
\begin{eqnarray}
\bf{S}(\lambda) &=& \left(\begin{array}{c}
S^{(1)}(\lambda)\\
S^{(2)}(\lambda)
\end{array}
\right),\quad \bf{T}(\lambda) = \left(\begin{array}{c}
T^{(1)}(\lambda)\\
T^{(2)}(\lambda)
\end{array}
\right),\\%
\bf{f_S}(\lambda) &=& \left(\begin{array}{c}
f_S^{(1)}(\lambda)\\
f_S^{(2)}(\lambda)
\end{array}
\right),\quad \bf{b_S}(\lambda) = \left(\begin{array}{c}
b_S^{(1)}(\lambda)\\
b_S^{(2)}(\lambda)
\end{array}
\right), \\
\bf{f_T}(\lambda) &=& \left(\begin{array}{c}
f_T^{(1)}(\lambda)\\
f_T^{(2)}(\lambda)
\end{array}
\right),\quad \bf{b_T}(\lambda) = \left(\begin{array}{c}
b_T^{(1)}(\lambda)\\
b_T^{(2)}(\lambda)
\end{array}
\right),\\%
\bf{K}(\lambda)&=&\left(\begin{array}{cc}
h(\lambda)&g(\lambda+\lambda^{(1)}_0-\lambda^{(2)}_0)\\
g(\lambda+\lambda^{(2)}_0-\lambda^{(1)}_0)&h(\lambda)
\end{array}\right).
\end{eqnarray}
We denote Fourier transforms of $\bf{S}(\lambda)$ and
$\bf{T}(\lambda)$ as
\begin{eqnarray}
\bf{\hat{S}}(\omega)&=&\int_{-\infty}^\infty d\lambda
e^{\mathrm{i}\lambda
\omega} \bf{S}(\lambda),\\
\bf{\hat{T}}(\omega)&=&\int_{-\infty}^\infty d\lambda
e^{\mathrm{i}\lambda \omega} \bf{T}(\lambda),
\end{eqnarray}
which represents a Fourier transformation on each of the matrix
components.

Thus the Wiener-Hopf equations for $\bf{S}(\lambda)$ and
$\bf{T}(\lambda)$ in Fourier space are
\begin{eqnarray}
\bf{\hat{S}}(\omega)&=&\bf{\hat{f}_S}(\omega)+\bf{\hat{b}_S}(\omega)+\bf{\hat{K}}(\omega)\bf{\hat{S}}(\omega),\label{WHS}\\
\bf{\hat{T}}(\omega)&=&\bf{\hat{f}_T}(\omega)+\bf{\hat{b}_T}(\omega)
+\bf{\hat{K}}(\omega)\bf{\hat{T}}(\omega),\label{WHT}
\end{eqnarray}
where
\begin{eqnarray}
\bf{\hat{K}}(\omega) &=& \left(\begin{array}{cc}
\hat{h}(\omega)&e^{-\mathrm{i}\omega(\lambda^{(1)}_0-\lambda^{(2)}_0)}\hat{g}(\omega)\\
e^{\mathrm{i}\omega(\lambda^{(1)}_0-\lambda^{(2)}_0)}\hat{g}(\omega)&\hat{h}(\omega)
\end{array}\right).
\end{eqnarray}
Notice that when $\omega$ and $\lambda^{(r)}_0$ are real, the kernel
$\bf{\hat{K}}(\omega)$ is Hermitian, i.e.
\begin{equation}
\bf{\hat{K}^\dag}(\omega)=\bf{\hat{K}}(\omega).\label{hermite}
\end{equation}
Equations (\ref{WHS}) and (\ref{WHT}) can also be written as
\begin{eqnarray}
\bf{\hat{S}}(\omega)&=&\left(\bf{I}-\bf{\hat{K}}(\omega)\right)^{-1}\left(\bf{\hat{f}_S}(\omega)+\bf{\hat{b}_S}(\omega)\right),\label{hats}\\
\bf{\hat{T}}(\omega)&=&\left(\bf{I}-\bf{\hat{K}}(\omega)\right)^{-1}\left(\bf{\hat{f}_T}(\omega)
+\bf{\hat{b}_T}(\omega)\right).\label{hatt}
\end{eqnarray}

%The function $\left(1-\bf{\hat{K}}(\omega)\right)^{-1}$ can be
%simplified as
%\begin{eqnarray}
%\left(1-\bf{\hat{K}}(\omega)\right)^{-1} &=&
%\frac{1}{1-\hat{s}~\hat{a}_1} \left(\begin{array}{ll}
%1 &-\hat{s}~ e^{-i\omega(\lambda^{(1)}_0-\lambda^{(2)}_0)}\\
%-\hat{s}~e^{i\omega(\lambda^{(1)}_0-\lambda^{(2)}_0)}&1
%\end{array}\right).
%\end{eqnarray}

Since the function $\left(\bf{I}-\bf{\hat{K}}(\omega)\right)^{-1}$
is nonsingular, hermitian and positive definite at $\omega=0$, it is
also positive definite for $-\infty<\omega<\infty$. From Theorem 8.2
of Gohberg and Krein \cite{Gohberg}, the function can be factorized
as follows,
\begin{equation}
\left(\bf{I}-\bf{\hat{K}}(\omega)\right)^{-1}=\bf{G_+}(\omega)
\bf{G_-}(\omega),\label{factori}
\end{equation}
where each element of the $2\times 2$ matrix $\bf{G_+}(\omega)$
($\bf{G_-}(\omega)$) is analytic on the upper-half (lower-half)
plane. The limit $\bf{G_\pm}(\omega)$ is chosen to satisfy the
condition
\begin{equation}
\bf{G_\pm}(\infty)=\bf{I},
\end{equation}
where $\bf{I}$ is the identity matrix. From (\ref{hermite}), it
follows that
\begin{equation}
\bf{G_+}(-\omega) = \bf{G_-^T}(\omega), \label{Gswap}
\end{equation}
where $\bf{A}^{T}$ represents the transpose of the matrix $\bf{A}$.

In Appendix A and Appendix B, we have worked out the solutions to
equations (\ref{hats}) and (\ref{hatt}) in detail. We showed in
equations (\ref{Somega}) and (\ref{Tomega}) that their solutions are
\begin{equation}
\mathbf{\hat{S}}(\omega)= \mathrm{i}\frac{2\pi P}{\sqrt{3}c}
\left(\frac{1} {\omega+\mathrm{i}\epsilon} -\frac{1}{\omega+2\pi
\mathrm{i}/3c}\right)\mathbf{G_+}(\omega)\mathbf{G_-}(0)\left(
                                         \begin{array}{c}
                                           1 \\
                                           1 \\
                                         \end{array}
                                       \right)\label{Somega}
\end{equation}
and
\begin{equation}
\bf{\hat{T}}(\omega)=(\bf{G_{+}}(\omega)-\bf{I})\bf{V}
\label{Tomega}
\end{equation}
where $\bf{V}$ is a $2\times 1$ matrix given in equation (\ref{V}).

\section{Universal low temperature properties and spin-charge separation}
To solve the TBA equations (\ref{TBA}) and (\ref{TBA2}), we must
find a way to express them involving the functions $\hat{S}$ and
$\hat{T}$. Let us rewrite the dressed energy from the first set of
TBA equations (\ref{TBA}) as
\begin{eqnarray}
\nonumber \varepsilon\left(k\right) &=& k^{2}-\mu -2\pi P s_1\ast
a_1(k) - T\int_{-\infty}^\infty
s_1(k-\lambda)\ln(1+e^{\phi_1^{(1)}/T})d\lambda \\ && -
T\int_{-\infty}^\infty
s_2(k-\lambda)\ln(1+e^{\phi_1^{(2)}/T})d\lambda\nonumber\\
%&=& k^{2}-\mu -2\pi P s_1\ast a_1(k) -
%T\int_{|\lambda|>\lambda_{0}^{(1)}}
%s_1(k-\lambda)\ln(1+e^{\phi_1^{(1)}/T})d\lambda-
%T\int_{-\lambda_0^{(1)}}^{\lambda_0^{(1)}}
%s_1(k-\lambda)\ln(1+e^{\phi_1^{(1)}/T})d\lambda \nonumber\\
%&& -T\int_{|\lambda|>\lambda_{0}^{(2)}}
%s_2(k-\lambda)\ln(1+e^{\phi_1^{(2)}/T})d\lambda
%-T\int_{-\lambda_0^{(2)}}^{\lambda_0^{(2)}}
%s_2(k-\lambda)\ln(1+e^{\phi_1^{(2)}/T})d\lambda\nonumber\\
&\approx& \nonumber k^2-\mu -2\pi P s_1\ast a_1(k) -
T\int_{|\lambda|>\lambda_{0}^{(1)}}
s_1(k-\lambda)\ln(1+e^{\phi_1^{(1)}/T})d\lambda \\ && -
T\int_{-\lambda_0^{(1)}}^{\lambda_0^{(1)}}
s_1(k-\lambda)\ln(1+e^{\phi_1^{(1)}/T})d\lambda
-T\int_{|\lambda|>\lambda_{0}^{(2)}}
s_2(k-\lambda)\ln(1+e^{\phi_1^{(2)}/T})d\lambda
\nonumber\\
&& -T\int_{-\lambda_0^{(2)}}^{\lambda_0^{(2)}}
s_2(k-\lambda)\ln(1+e^{\phi_1^{(2)}/T})d\lambda\nonumber\\
&\approx& k^2-\mu -2\pi P s_1\ast a_1(k) -
T\int_{|\lambda|>\lambda_{0}^{(1)}}
s_1(k-\lambda)\ln(1+e^{\phi_1^{(1)}/T})d\lambda \nonumber\\
&& -T\int_{|\lambda|>\lambda_{0}^{(2)}}
s_2(k-\lambda)\ln(1+e^{\phi_1^{(2)}/T})d\lambda -\sum_{r=1}^2
\frac{\pi^2 T^2 \kappa^{(r)}}{3\sqrt{3}cS^{(r)\prime}(0)}
\end{eqnarray}
where we evaluated the the integration between the Fermi points as
\begin{eqnarray}
\nonumber
\sum_{r=1}^{2}T\int_{-\lambda_{0}^{(r)}}^{\lambda_{0}^{(r)}}s_{r}(k-\lambda)\ln(1+e^{\phi_{1}^{(r)}/T})d\lambda
&\approx& \nonumber
\sum_{r=1}^{2}T\int_{-\infty}^{\infty}s_{r}(k-\lambda)\ln(1+e^{-|\phi_{1}^{(r)}|/T})d\lambda
\\ \nonumber &\approx&
\sum_{r=1}^{2}\frac{\pi^{2}T^{2}}{6t^{(r)}}\left[s_{r}(k-\lambda_{0}^{(r)})+s_{r}(k+\lambda_{0}^{(r)})\right]
\\ \nonumber &\approx&
\sum_{r=1}^{2}\frac{\pi^{2}T^{2}}{3\sqrt{3}ct^{(r)}}e^{-2\pi\lambda_{0}^{(r)}/3c}
\\ &=& \sum_{r=1}^{2}
\frac{\pi^2 T^2 \kappa^{(r)}}{3\sqrt{3}cS^{(r)\prime}(0)}.
\end{eqnarray}
From the third line to the fourth line, we made use of the fact that
\begin{equation}
S^{(r)\prime}(0)=e^{2\pi\lambda_{0}^{(r)}/3c}t^{(r)}\kappa^{(r)}.
\end{equation}

The other two integrals can be simplified as
\begin{eqnarray}
&&\int_{\lambda_0^{(r)}}^{\infty} s_r(k-\lambda) [\phi^{(r)}
+\eta^{(r)}]d\lambda\nonumber\\
&& =\int_{0}^{\infty} s_r(k-\lambda-\lambda_0^{(r)})
\left[\frac{e^{-2\pi\lambda_0^{(r)}/3c}}
{\kappa^{(r)}}S^{(r)}(\lambda) + \frac{\pi^2 T^2 \kappa^{(r)}}
{6e^{-2\pi\lambda_0^{(r)}/3c}}
T^{(r)}(\lambda)\right] d\lambda\nonumber\\
&& \approx\int_{0}^{\infty} \frac{e^{2\pi(k-\lambda-
2\lambda_0^{(r)})/3c}} {\sqrt{3}c\kappa^{(r)}}S^{(r)}(\lambda)
d\lambda + \int_{0}^{\infty} \frac{\pi^2 T^2
\kappa^{(r)}e^{2\pi(k-\lambda)/3c}} {6\sqrt{3}c} T^{(r)}(\lambda) d\lambda\nonumber\\
&& =\int_{0}^{\infty} \frac{\sqrt{3} c H^2 \kappa^{(r)}} {4\pi^2
P^2} e^{2\pi(k-\lambda)/3c} S^{(r)}(\lambda) d\lambda +
\int_{0}^{\infty} \frac{\pi^2 T^2
\kappa^{(r)}e^{2\pi(k-\lambda)/3c}} {6\sqrt{3}c}
T^{(r)}(\lambda) d\lambda\nonumber\\
&& \nonumber =\int_{0}^{\infty}\int_{-\infty}^{\infty}
\frac{\sqrt{3} c H^2 \kappa^{(r)}} {8\pi^3 P^2}
e^{2\pi(k-\lambda)/3c}e^{-\mathrm{i}\omega\lambda}
\hat{S}^{(r)}(\omega) d\omega d\lambda \\ && \quad
+\int_{0}^{\infty} \int_{-\infty}^{\infty} \frac{\pi T^2
\kappa^{(r)}} {12\sqrt{3}c} e^{2\pi(k-\lambda)/3c}
e^{-\mathrm{i}\omega\lambda}
\hat{T}^{(r)}(\omega) d\omega d\lambda\nonumber\\
&& \approx\int_{-\infty}^{\infty} \frac{\sqrt{3} c H^2 \kappa^{(r)}}
{8\pi^3
P^2}\frac{\hat{S}^{(r)}(\omega)}{\mathrm{i}\omega+2\pi/3c}e^{2\pi
k/3c}d\omega +\int_{-\infty}^{\infty} \frac{\pi T^2
\kappa^{(r)}}{12\sqrt{3}c}
\frac{\hat{T}^{(r)}(\omega)}{\mathrm{i}\omega+2\pi/3c}e^{2\pi k/3c}d\omega\nonumber\\
&& =\frac{\sqrt{3} c H^2 \kappa^{(r)}} {4\pi^2 P^2}
\hat{S}^{(r)}\left(\frac{2\pi \mathrm{i}}{3 c}\right) +\frac{\pi^2
T^2 \kappa^{(r)}} {6\sqrt{3}c} \hat{T}^{(r)}\left(\frac{2\pi
\mathrm{i}}{3 c}\right).
\end{eqnarray}
Thus the dressed energy is
\begin{eqnarray}
\varepsilon\left(k\right) &\approx& k^{2}-\mu -2\pi P s_1\ast a_1(k)
- 2\sum_{r=1}^{2}\int_{\lambda_0^{(r)}}^{\infty} s_r(k-\lambda)
[\phi^{(r)} +\eta^{(r)}]d\lambda-\sum_{r=1}^2
\frac{\pi^2T^2 \kappa^{(r)}} {3\sqrt{3}c S^{(r)\prime}(0)}\nonumber\\
&=& \nonumber k^2-\mu -2\pi P s_1\ast a_1(k) \\ && -\sum_{r=1}^2
\left[\frac{\sqrt{3} c H^2 \kappa^{(r)}} {2\pi^2 P^2}
\hat{S}^{(r)}\left(\frac{2\pi i}{3 c}\right) +\frac{\pi^2 T^2
\kappa^{(r)}} {3\sqrt{3}c} \hat{T}^{(r)}\left(\frac{2\pi i}{3
c}\right)+\frac{\pi^2T^2 \kappa^{(r)}}
{3\sqrt{3}cS^{(r)\prime}(0)}\right].
\end{eqnarray}

From (\ref{Somega}),
$\mathbf{\hat{S}}\left({2\pi\mathrm{i}}/{3c}\right)$ is expressed as
\begin{eqnarray}
\mathbf{\hat{S}}\left(\frac{2\pi\mathrm{i}}{3c}\right)=\frac{\sqrt{3}}{2}P
\mathbf{G_+}\left(\frac{2\pi
\mathrm{i}}{3c}\right)\mathbf{G_-}(0)\left(
                                                     \begin{array}{c}
                                                       1 \\
                                                       1 \\
                                                     \end{array}
                                                   \right).
\end{eqnarray}
To evaluate the sum $\sum_{r=1}^{2}\kappa^{(r)}\hat{S}^{(r)}(2\pi
\mathrm{i}/3c)$, we only need to consider the matrix product
\begin{eqnarray}
\left(
  \begin{array}{cc}
    \kappa^{(1)} & \kappa^{(2)} \\
  \end{array}
\right) \mathbf{\hat{S}}\left(\frac{2\pi \mathrm{i}}{3c}\right) &=&
\frac{\sqrt{3}}{2} P \left(
  \begin{array}{cc}
    \kappa^{(1)} & \kappa^{(2)} \\
  \end{array}
\right)\mathbf{G_+}\left(\frac{2\pi \mathrm{i}}{3c}\right)
\mathbf{G_-}\left(0\right)\left(
                        \begin{array}{c}
                          1 \\
                          1 \\
                        \end{array}
                      \right)
\nonumber\\
&=& \frac{\sqrt{3}}{2} P \left(
                           \begin{array}{cc}
                             1 & 1 \\
                           \end{array}
                         \right)
\mathbf{G_+}\left(0\right)\mathbf{G_-}\left(0\right)\left(
                        \begin{array}{c}
                          1 \\
                          1 \\
                        \end{array}
                      \right)
\nonumber\\
&=& \sqrt{3} P.
\end{eqnarray}
We made use of equations (\ref{Gswap}) and (\ref{kappa1}) to derive
the second line from the first line. The term
$\mathbf{G_+}(0)\mathbf{G_-}(0)$ was explicitly derived from its
original definition (\ref{factori}). To do that, we needed to find
$\mathbf{\hat{K}}(0)$. Since
\begin{equation}
\hat{s}_{1}(0)=\int_{-\infty}^{\infty}s_{1}(x)dx=\frac{2}{3},\qquad
\hat{s}_{2}(0)=\int_{-\infty}^{\infty}s_{2}(x)dx=\frac{1}{3},
\end{equation}
hence
\begin{equation}
\mathbf{G_+}(0)\mathbf{G_-}(0)=\left(\mathbf{I}-\mathbf{\hat{K}}(0)\right)^{-1}=\left(
                                                                        \begin{array}{cc}
                                                                          2 & -1 \\
                                                                          -1 & 2 \\
                                                                        \end{array}
                                                                      \right).
\end{equation}

%The next step is to evaluate the expression $\kappa\left(\hat{T}
%\left(2\pi i/3c\right) + V\right)$. From (\ref{Tomega}), we have
%\begin{eqnarray}
%\hat{T}=G_+ P_+\left(G_-\hat{f}(\omega)\right)=
%\left({G_+}-1\right)V.
%\end{eqnarray}

The next step is to evaluate the
$\sum_{r=1}^{2}\kappa^{(r)}\left[\hat{T}^{(r)}(2\pi
\mathrm{i}/3c)+1/S^{(r)\prime}(0)\right]$. Similar to the previous
calculation, it is easier to work with the matrix representations.
Doing so yields
\begin{eqnarray}
\nonumber \left(
            \begin{array}{cc}
              \kappa^{(1)} & \kappa^{(2)} \\
            \end{array}
          \right)\left(\mathbf{\hat{T}}\left(\frac{2\pi
          \mathrm{i}}{3c}\right)+\mathbf{V}\right) &=& \left(
            \begin{array}{cc}
              \kappa^{(1)} & \kappa^{(2)} \\
            \end{array}
          \right)\mathbf{G_+}\left(\frac{2\pi
          \mathrm{i}}{3c}\right)\mathbf{V} \\ \nonumber &=& \left(
                                                        \begin{array}{cc}
                                                          1 & 1 \\
                                                        \end{array}
                                                      \right)\mathbf{G_+}(0)\mathbf{V}
                                                      \\ \nonumber
                                                      &=&
                                                      \frac{3\sqrt{3}c^{2}}{4\pi^{2}P}\mathbf{S'^{T}}(0)\mathbf{V}
                                                      \\ &=&
                                                      \frac{3\sqrt{3}c^{2}}{2\pi^{2}P}
\end{eqnarray}
where we made use of the transpose of equation (\ref{Sprime}), which
is
\begin{equation}
\mathbf{S'^{T}}(0)=\frac{4\pi^{2}P}{3\sqrt{3}c^{2}}\left(
                                                 \begin{array}{cc}
                                                   1 & 1 \\
                                                 \end{array}
                                               \right)\mathbf{G_+}(0).
\end{equation}

%Therefore we have
%\begin{eqnarray}
%\kappa\left(\hat{T}\left(\frac{2\pi
%i}{3c}\right)+V\right)=\frac{3\sqrt{3}c^2}{4\pi^2 P}.
%\end{eqnarray}

Finally, the dressed energy can be written as
\begin{equation}
\varepsilon(k)\approx k^{2}-\mu -2\pi P s_1\ast a_1(k) -
\frac{3cH^2}{2\pi^2 P}-\frac{cT^2}{2 P}. \label{dressedEnergy}
\end{equation}
Here
\begin{eqnarray}
\nonumber s_1\ast a_1(k) &=&
\int_{-\infty}^{\infty}s_1(k-\lambda)a_1(\lambda)d\lambda \\
\nonumber &\approx&
\int_{-\infty}^{\infty}s_1(\lambda)a_1(\lambda)d\lambda \\ &=& \frac
1{6\sqrt{3} c}+\frac{\ln 3}{2\pi c}.
\end{eqnarray}
Through integration by parts, equation (\ref{pressure}) is given by
\begin{equation}
P=\frac{1}{\pi}\int_{0}^{\infty }\frac{\sqrt{ \varepsilon}
d\varepsilon}{1+e^{{(\varepsilon -A)}/{T}}}. \label{pressure2}
\end{equation}
where the dressed energy is rewritten in the form
\begin{equation}
\varepsilon(k)= k^{2}-A(T,H)
\end{equation}
with
\begin{equation}
A(T,H)=\mu+2\pi P s_1\ast a_1(k)+\frac{3cH^2}{2 \pi^2
P}+\frac{cT^2}{2 P}.
\end{equation}

For finite temperature, the pressure can be given by polylogarithm
function
\begin{equation}
P=-\sqrt{\frac{1}{4\pi}}T^{ \frac{3}{2}} \mathrm{{Li}}
_{\frac{3}{2}}(-e^{\frac{A}{T}})\label{EOS}
\end{equation}
This equation describes the exact low temperature thermodynamics and
thus the full phase diagram can be given in finite temperature in a
much larger regime than that given by field theory. However, the low
temperature properties cannot be seen directly from the polylog
functions. Therefore, we apply Sommerfeld's expansion technique for
this equation and study the universal low temperature properties.
The expansion of the pressure is in terms of powers of
$\left(T/A(T,H)\right)$. Consider the leading order, the equation is
\begin{equation}
P=\frac{2}{3\pi} A^{\frac{3}{2}}\left[ 1+\frac{\pi
^{2}}{8}\left(\frac{T}{A}\right)^{2}\right].  \label{p_gamma}
\end{equation}
Furthermore, with the relation $n=\partial P/\partial \mu$, after
the iteration, and neglecting higher order terms of $T^2$, $H^2$ and
$1/c$, the chemical potential $\mu$ is obtained as
\begin{eqnarray}
\nonumber \mu &=& n^2\pi^2 \left[ 1 -\frac{16\pi n s_1\ast
a_1(k)}{3}+\frac{9 c H^2}{4 n^5 \pi^6}\left(1+3\pi n s_1 \ast
a_1(k)\right)\right. \\ && \left.+\frac{3 c T^2}{2 n^5
\pi^4}\left(1+3\pi n s_1 \ast a_1(k)\right) + \frac{T^2}{12 n^4
\pi^2}\right]. \label{mu}
\end{eqnarray}
Substituting (\ref{mu}) into (\ref{p_gamma}), the pressure $P$
becomes
\begin{eqnarray}
\nonumber P &=& \frac{2}{3\pi} n^3 \pi^3 \left[ 1 - 6\pi n s_1\ast
a_1(k) + \frac{81 c H^2}{8 n^5 \pi^6}\left(1+4\pi n s_1 \ast
a_1(k)\right)\right. \\ && \left.+\frac{27 c T^2}{8 n^5
\pi^4}\left(1+4\pi n s_1 \ast a_1(k)\right) + \frac{T^2}{4 n^4
\pi^2}\left(1+2 \pi n s_1\ast a_1(k)\right)\right]. \label{press}
\end{eqnarray}
Finally, with the relation $F=\mu n - P$, the free energy suggests a
universal low temperature behavior of TLL, namely
\begin{eqnarray}
F&=&\mu n - P\nonumber\\
&=& \nonumber \frac{1}{3} n^3 \pi^2\left(1-4\pi n s_1\ast
a_1(k)\right)
-\frac{9 c H^2}{4 n^2 \pi^4}\left(1+6\pi n s_1 \ast a_1(k)\right) \\
&& -\frac{3 c T^2}{4 n^2 \pi^2}\left(1+6\pi n s_1 \ast a_1(k)\right)
- \frac{T^2}{12
n}\left(1 + 4 \pi n s_1\ast a_1(k)\right)\nonumber\\
&=&E_0 -\frac{\pi
T^2}{6}\left(\frac{C_s}{v_s}+\frac{C_c}{v_c}\right). \label{F}
\end{eqnarray}
where
\begin{eqnarray}
E_0 &=& \frac{1}{3} n^3 \pi^2\left(1-4\pi n s_1\ast a_1(k)\right)
- \frac{9 c H^2}{4 n^2 \pi^4}\left(1+6\pi n s_1 \ast a_1(k)\right)\nonumber\\
&=& \frac{1}{3} n^3 \pi^2\left(1-\frac{2\pi n}{3\sqrt{3}c}-
\frac{2n\ln3}{c}\right) -\frac{9 c H^2}{4 n^2
\pi^4}\left(1+\frac{\pi n}{\sqrt{3}c}+\frac{3n\ln3}{c}\right)\\
v_s&=&\frac{4}{9c}n^2\pi^3 \left(1-6\pi n s_1 \ast
a_1(k)\right)\nonumber\\
&=&\frac{4}{9c}n^2\pi^3\left(1-\frac{\pi n}{\sqrt{3}c}-\frac{3n\ln3}{c} \right)\\
v_c&=& 2n\pi \left(1-4\pi n s_1\ast a_1(k)\right)\nonumber\\
&=& 2n\pi \left(1-\frac{2n\pi}{3\sqrt{3}c}-\frac{2n\ln3}{c}\right).
\end{eqnarray}

The spin and charge velocities can be derived from the relations
$v_c = \varepsilon^\prime\left(k_0\right)/2\pi
\rho_c\left(k_0\right)$ and $v_s =
\phi_1^{(r)\prime}\left(\lambda_0\right)/2\pi
\rho_s\left(\lambda_0\right)$ \cite{Frahm}. For three-component
fermions, there are two spin velocities $v_{s1}$ and $v_{s2}$, where
$v_{s1}=v_{s2}$. The central charge for the spin part is $C_s=2$ and
that for the charge part is $C_c=1$. The reason for $C_s=2$ is
because the $SU(3)$ invariant fermion model has two spin ``Fermi
seas'' whose dependence on $H$ are equal i.e., we considered the
case where $H_{1}=H_{2}=H$. This result shows that spin-charge
separation exists for low temperatures and a small external magnetic
field. This result is coincident with the results from conformal
field theory \cite{Affleck, Blote}.

\begin{figure}
\centering
\begin{tabular}{cc}
\epsfig{file=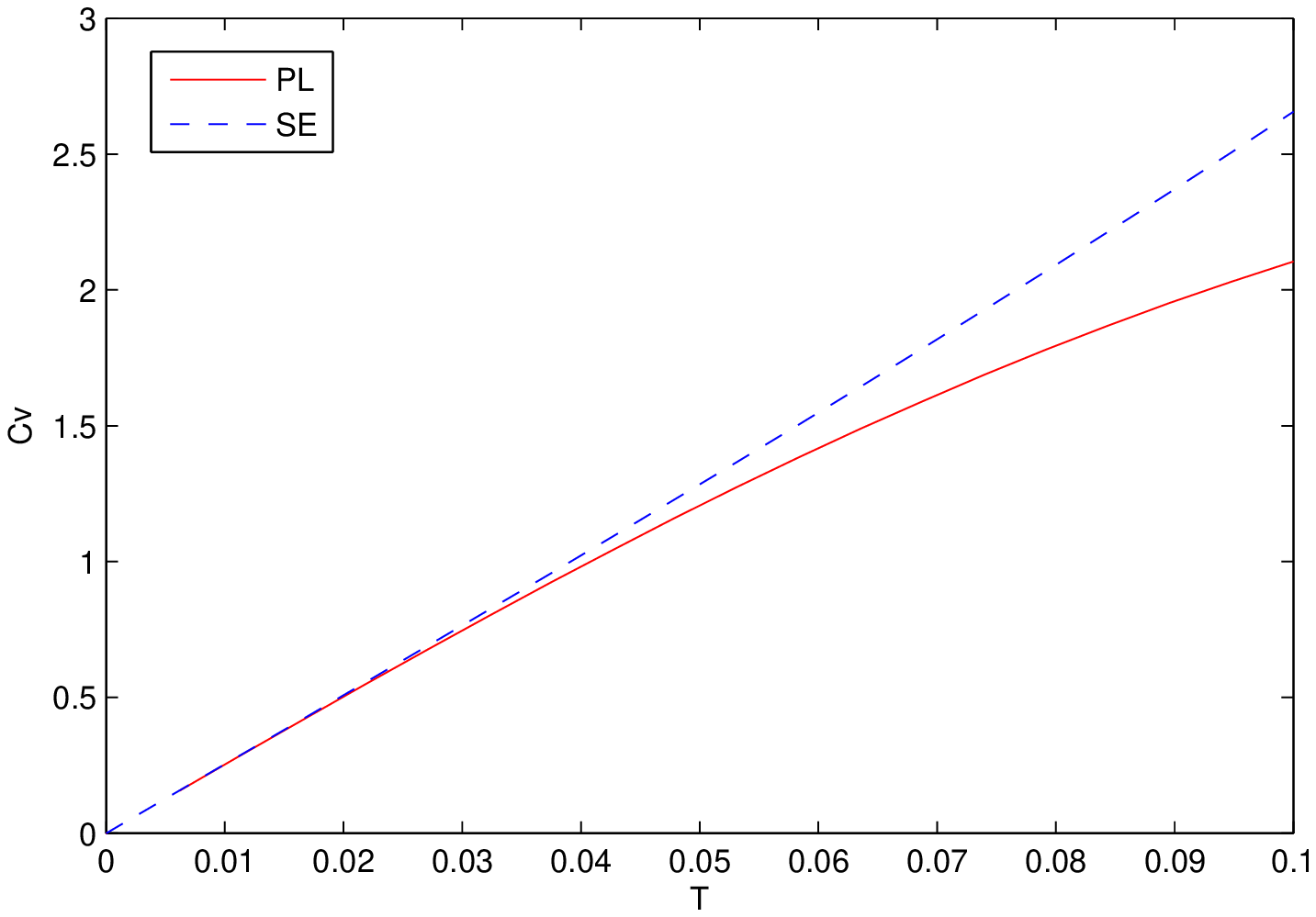,width=0.5\linewidth,clip=} &
\epsfig{file=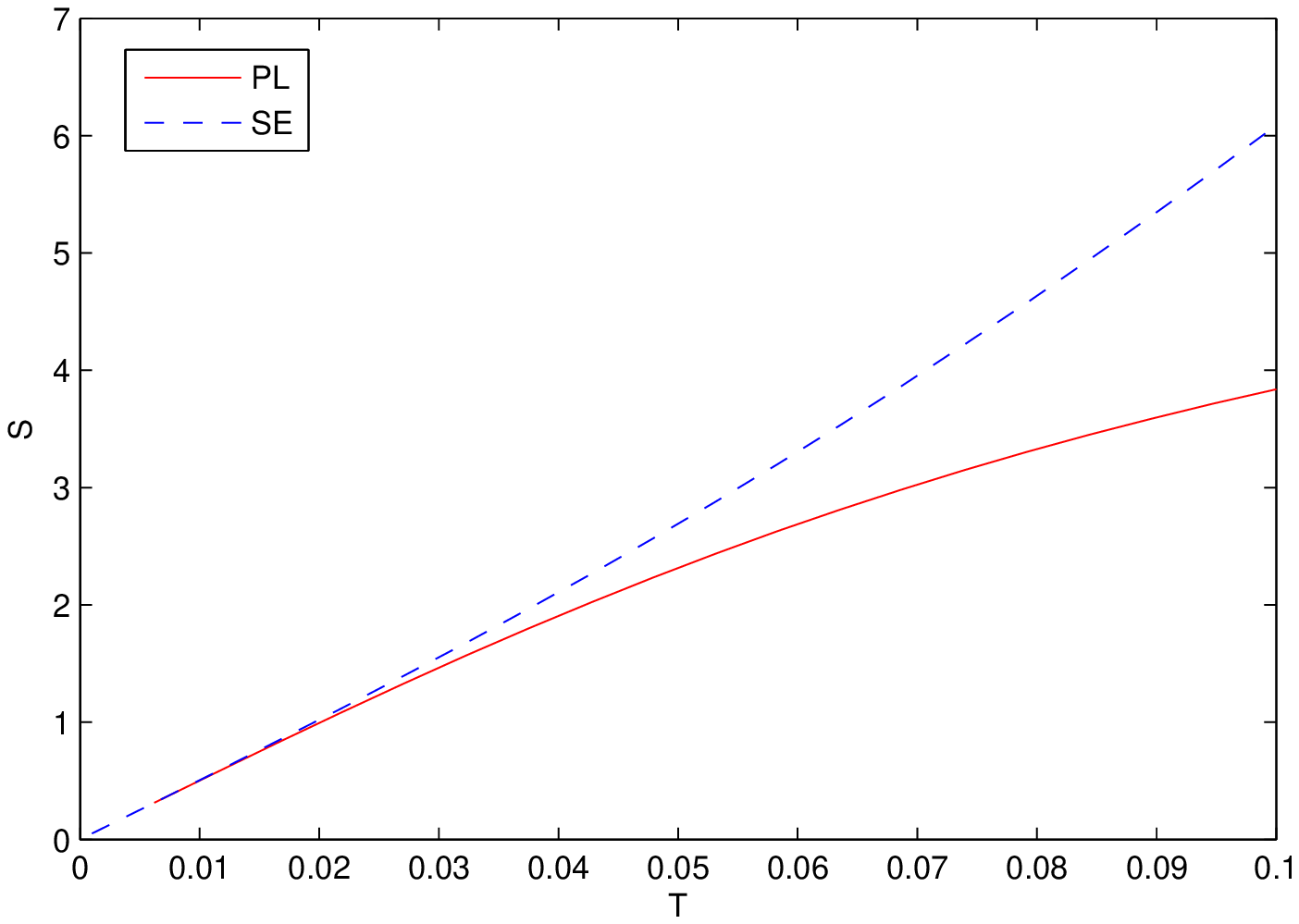,width=0.5\linewidth,clip=}
\end{tabular}\caption{(Color online) The left plot shows
$C_V-T$ curves for external magnetic field $H=0.001$, chemical
potential $\mu=6$ and coupling constant $c=30$. The right plot shows
$S-T$ curves for external magnetic field $H=0.001$, chemical
potential $\mu=3$ and coupling constant $c=50$. See text}
\label{fig:cv-s-t}
\end{figure}

To compare the results from Sommerfeld's expansion and the polylog
function, we plot graphs of specific heat $C_V-T$ and entropy
$S_V-T$ versus temperature in FIG.~\ref{fig:cv-s-t}. Both graphs
show that at low temperatures in the strong coupling regime,
Sommerfeld's expansion agrees well with the polylog function.
However, when the temperature increases, the deviation between the
two curves become more apparent. This means that the
Tomonaga-Luttinger liquid form of the free energy is only valid at
very low temperatures and conformal invariance breaks down as the
excitations take place at higher temperatures.

%\begin{figure}[t]
%{{\includegraphics [width=0.7\linewidth]{Cv_T.eps}}} \caption{The
%$C_V-T$ curves for external magnetic field $H=0.001$, chemical
%potential $\mu=6$ and coupling constant $c=30$. } \label{fig:cv-t}
%\end{figure}

%\begin{figure}[t]
%{{\includegraphics [width=0.7\linewidth]{S_T.eps}}} \caption{The
%S-T$ curves for external magnetic field$H=0.001$, chemical potential
%$\mu=3$ and coupling constant $c=50$.}
%\label{fig:s-t}
%\end{figure}

\section{Conclusion}

We have presented a systematic  way to study low temperature
behavior of systems with high spin symmetry and with repulsive
interactions  via the Wiener-Hopf method. In particular, we have
derived the universal thermodynamics of 1D strongly repulsive
fermions with $SU(3)$ symmetry under a weak external magnetic field.
We have applied the Wiener-Hopf method to obtain the universal low
energy physics in terms of spin-charge separation. The chemical
potential (\ref{mu}), pressure (\ref{press}) and free energy
(\ref{F}) of the system have been derived analytically at low
temperatures. The free energy gives insight into understanding the
spin-charge separation theory for the higher spin systems. Moreover,
beyond the spin-charge separation regime, the thermodynamics
(\ref{EOS}) has been given in terms of the polylog function. This
result  can be used to test universal scaling behavior in the
quantum critical regime. Our results open up further study of
quantum criticality in 1D interacting Fermi gases with higher spin
symmetry and repulsive interaction.

\ack

This work is in part supported by the NSFC, the Knowledge Innovation
Project of Chinese Academy of Sciences, the National Program for
Basic Research of MOST(China) and the Australian Research Council.

\clearpage

\appendix
\section{Solution to Equation (\ref{hats})}
With the product decomposition given in (\ref{factori}), equation
(\ref{hats}) can be written as
\begin{equation}
\mathbf{G}_+^{-1}(\omega)\mathbf{\hat{S}}(\omega)=\mathbf{G_-}(\omega)\mathbf{\hat{f}_S}(\omega)
+\mathbf{G_-}(\omega)\mathbf{\hat{b}_S}(\omega). \label{gpm}
\end{equation}
The function $\mathbf{G}_+^{-1}(\omega)\mathbf{\hat{S}}(\omega)$ is
analytic on the upper-half plane by the definitions of
$\mathbf{G}_+^{-1}(\omega)$ and $\mathbf{\hat{S}}(\omega)$. On the
other hand, $\mathbf{G_-}(\omega)\mathbf{\hat{b}_S}(\omega)$ is
analytic on the lower-half plane. We are left with the function
$\mathbf{G_-}(\omega)\mathbf{\hat{f}_S}(\omega)$ which we have to
decompose into the sum of a function which is analytic on the
upper-half plane and another function which is analytic on the
lower-half plane.

First, notice that $\hat{f}^{(r)}_S(\omega)$ can be written as
\begin{equation}
\hat{f}_S^{(r)}(\omega)=\frac{2\pi P
\mathrm{i}}{\sqrt{3}c}\left(\frac{1}{\omega+\mathrm{i}\epsilon}-\frac{\kappa^{(r)}}{\omega+2\pi
\mathrm{i}/3c}\right)
\end{equation}
where $\epsilon$ is small. The function
$\mathbf{G_-}(\omega)\mathbf{\hat{f}_S}(\omega)$ in matrix form is
then decomposed into
\begin{eqnarray}
\nonumber \mathbf{G_-}(\omega)\mathbf{\hat{f}_S}(\omega) &=&
\frac{2\pi P
\mathrm{i}}{\sqrt{3}c}\left[\frac{1}{\omega+\mathrm{i}\epsilon}\left(\mathbf{G_-}(\omega)-\mathbf{G_-}(-\mathrm{i}\epsilon)\right)\left(
                                                                                                 \begin{array}{c}
                                                                                                   1 \\
                                                                                                   1 \\
                                                                                                 \end{array}
                                                                                               \right)\right.
\\ && \left.-\frac{1}{\omega+2\pi\mathrm{i}/3c}\left(\mathbf{G_-}(\omega)-\mathbf{G_-}(-2\pi
\mathrm{i}/3c)\right)\left(\begin{array}{l}
\kappa^{(1)}\\
\kappa^{(2)}
\end{array}
\right)\right]\nonumber\\
&& +\frac{2\pi P\mathrm{i}}{\sqrt{3}c} \left[\frac{1}
{\omega+\mathrm{i}\epsilon}\mathbf{G_-}(-\mathrm{i}\epsilon)\left(
                                        \begin{array}{c}
                                          1 \\
                                          1 \\
                                        \end{array}
                                      \right)
-\frac{1}{\omega+2\pi i/3c}\mathbf{G_-}(-2\pi
\mathrm{i}/3c)\left(\begin{array}{l}
\kappa^{(1)}\\
\kappa^{(2)}
\end{array}
\right)\right]\nonumber\\
&\equiv&\mathbf{\Phi^-}(\omega)+\mathbf{\Phi^+}(\omega).
\end{eqnarray}

Thus (\ref{gpm}) can be rewritten as
\begin{equation}
\mathbf{G}_+^{-1}(\omega)\mathbf{\hat{S}}(\omega)-\mathbf{\Phi^+}(\omega)
=\mathbf{\Phi^-}(\omega)+\mathbf{G_-}(\omega)\mathbf{\hat{b}_S}(\omega)
\end{equation}
where the left-hand side is analytic on the upper-half plane, and
right-hand side is analytic on the lower-half plane. When $\omega
\rightarrow \infty$, both the left-hand side and right-hand side
tend to zero, thus
\begin{equation}
\mathbf{G}_+^{-1}(\omega)\mathbf{\hat{S}}(\omega)-\mathbf{\Phi^+}(\omega)=0
\end{equation}
which gives
\begin{eqnarray}
\nonumber \mathbf{\hat{S}}(\omega) &=&
\mathbf{G_+}(\omega)\mathbf{\Phi^+}(\omega) = \frac{2\pi P
\mathrm{i}}{\sqrt{3}c} \left[\frac{1}
{\omega+\mathrm{i}\epsilon}\mathbf{G_+}(\omega)\mathbf{G_-}(0)\left(
                                     \begin{array}{c}
                                       1 \\
                                       1 \\
                                     \end{array}
                                   \right)\right.
\\ && \left.-\frac{1}{\omega+2\pi\mathrm{i}/3c}\mathbf{G_+}(\omega)\mathbf{G_-}(-2\pi
\mathrm{i}/3c)\left(\begin{array}{l}
\kappa^{(1)}\\
\kappa^{(2)}
\end{array}
\right)\right].
\end{eqnarray}

Since $S^{(r)}(0)=0$ and $\hat{S}^{(r)}(\omega)$ is analytic on the
upper-half plane, the following result
\begin{equation}
\lim_{|\omega|\rightarrow\infty}\omega\mathbf{\hat{S}}(\omega)=0
\end{equation}
is given by contour integration. This boundary condition is
equivalent to
\begin{equation}
\mathbf{G_-}(-2\pi\mathrm{i}/3c)\left(\begin{array}{l}
\kappa^{(1)}\\
\kappa^{(2)}
\end{array}
\right)=\mathbf{G_-}(0)\left(
                                              \begin{array}{c}
                                                1 \\
                                                1 \\
                                              \end{array}
                                            \right).
\label{kappa1}
\end{equation}
Using this result we can show that
\begin{equation}
\mathbf{\hat{S}}(\omega)= \mathrm{i}\frac{2\pi P}{\sqrt{3}c}
\left(\frac{1} {\omega+\mathrm{i}\epsilon} -\frac{1}{\omega+2\pi
\mathrm{i}/3c}\right)\mathbf{G_+}(\omega)\mathbf{G_-}(0)\left(
                                         \begin{array}{c}
                                           1 \\
                                           1 \\
                                         \end{array}
                                       \right)
.\label{Somega}
\end{equation}

Besides, note that
\begin{equation}
\mathbf{{S}^\prime}(0)=
\frac{d}{d\lambda}\mathbf{S}(\lambda)\Big|_{\lambda=0^+} = -
\lim_{|\omega|\rightarrow\infty}\omega^2\mathbf{\hat{S}}(\omega),
\end{equation}
thus we find the expression of $\mathbf{S'}(0)$
\begin{equation}
\mathbf{S'}(0)=\frac{4\pi^2 P}{3\sqrt{3}c^2}\mathbf{G_-}(0)\left(
                                                           \begin{array}{c}
                                                             1 \\
                                                             1 \\
                                                           \end{array}
                                                         \right).
                                                         \label{Sprime}
\end{equation}

\section{Solution to Equation (\ref{hatt})}
Similarly for $\mathbf{\hat{T}}(\omega)$ with the factorization
(\ref{factori}), we find that
\begin{equation}
\mathbf{G}_+^{-1}(\omega)\mathbf{\hat{T}}(\omega)=\mathbf{G_-}(\omega)\mathbf{\hat{f}_T}(\omega)
+\mathbf{G_-}(\omega)\mathbf{\hat{b}_T}(\omega).\label{gpm2}
\end{equation}
With similar arguments as before, we only need to find a
decomposition of $\mathbf{G_-}(\omega)\mathbf{\hat{f}_T}(\omega)$
into the sum of two functions that are analytic on the upper-half
plane and lower-half plane, respectively. Consider the functions
\begin{eqnarray}
f^{(1)}_{T}(\lambda)&=& \frac{h(\lambda)}{S^{(1)\prime}(0)} +
\frac{g(\lambda+\lambda^{(1)}_0-\lambda^{(2)}_0)}{S^{(2)\prime}(0)},\\
f^{(2)}_{T}(\lambda)&=& \frac{h(\lambda)}{S^{(2)\prime}(0)} +
\frac{g(\lambda+\lambda^{(1)}_0-\lambda^{(2)}_0)}{S^{(1)\prime}(0)}.
\end{eqnarray}

After taking a Fourier transformation, the matrix
$\mathbf{\hat{f}_{T}}(\omega)$ can be expressed in terms of the
kernel $\mathbf{\hat{K}}(\omega)$ as
\begin{equation}
\mathbf{\hat{f}_{T}}(\omega)=\mathbf{\hat{K}}(\omega)\mathbf{V}
=\left(\mathbf{I}-\mathbf{G}_-^{-1}(\omega)\mathbf{G}_+^{-1}(\omega)\right)\mathbf{V},
\end{equation}
where
\begin{equation}
\mathbf{V} = \left(\begin{array}{l}
1/S^{(1)\prime}(0)\\
1/S^{(2)\prime}(0)
\end{array}
\right). \label{V}
\end{equation}
Multiplying $\mathbf{G_-}(\omega)$ on both sides of this equation,
we have
\begin{equation}
\mathbf{G_-}(\omega)\mathbf{\hat{f}_{T}}(\omega)=\left(\mathbf{G_-}(\omega)
-\mathbf{G}_+^{-1}(\omega)\right)\mathbf{V}.
\end{equation}
Substituting this relation back into equation (\ref{gpm2}) gives
\begin{equation}
\mathbf{G}_+^{-1}(\omega)\mathbf{\hat{T}}(\omega)+\mathbf{G}_+^{-1}(\omega)\mathbf{V}
=\mathbf{G_-}(\omega)\mathbf{V}+\mathbf{G_-}(\omega)\mathbf{\hat{b}_T}(\omega).
\end{equation}

Based on analyticity arguments, matrix elements on both sides of the
equation are equal to a function that we shall denote in matrix form
by $\mathbf{Q}(\omega)$, i.e.,
\begin{equation}
\mathbf{G}_+^{-1}(\omega)\mathbf{\hat{T}}(\omega)+\mathbf{G}_+^{-1}(\omega)\mathbf{V}=\mathbf{Q}(\omega).
\end{equation}
Similarly to $\mathbf{\hat{S}}(\omega)$, the function
$\mathbf{\hat{T}}(\omega)$ vanishes when $\omega\rightarrow\infty$.
Because $\mathbf{Q}(\omega)$ is an entire function that is bounded,
it is equal to a constant by Louville's Theorem. This constant can
be determined by taking the limit $\omega\rightarrow\infty$ which
gives $\mathbf{Q}=\mathbf{V}$. Hence we finally obtain an equation
for $\mathbf{\hat{T}}(\omega)$ expressed in terms of functions whose
values are known, i.e.,
\begin{equation}
\bf{\hat{T}}(\omega)=(\bf{G_{+}}(\omega)-\bf{I})\bf{V}.
\label{Tomega}
\end{equation}

\end{document}